\documentclass[aps,pra,floatfix,showpacs]{revtex4}
\usepackage{graphicx}
\usepackage{epsfig}
\usepackage{pst-plot}
\usepackage{bm}
\usepackage{longtable}
\begin{document}
 
\title{Vortices in Atomic Bose-Einstein Condensates in the Large Gas
  Parameter Region} 

\author{J.~K.~Nilsen$^{1}$, J.~Mur-Petit$^{2}$,
  M.~Guilleumas$^{2}$, M.~Hjorth-Jensen$^{1,3,4,5}$, 
  and A.~Polls$^{2}$}
\affiliation{ 
  $^1$ Department of Physics, University of Oslo, 
  N-0316 Oslo, Norway \\
  $^2$ Departament d'Estructura i Constituents de la Mat\`eria,
  Universitat de Barcelona, E-08028 Barcelona\\
  $^3$ Center of Mathematics for Applications, University of Oslo,
  N-0316 Oslo, Norway\\
$^4$PH Division, CERN, CH-1211 Geneve 23, Switzerland\\
$^5$Department of Physics and Astronomy,
Michigan State University, East Lansing, MI 48824, USA}
% \date{\empty}
\begin{abstract}
In this work we compare the results of the Gross-Pitaevskii and modified 
Gross-Pitaevskii
equations
with ab initio variational  Monte Carlo calculations for
Bose-Einstein condensates of atoms in axially symmetric  traps. We examine both the 
ground state and excited states having a vortex line along
the $z$-axis at high values of the gas parameter and
demonstrate an excellent agreement between the modified Gross-Pitaevskii
and ab initio Monte Carlo methods, both for the ground  and vortex states.
\end{abstract}
 
\maketitle
\section{Introduction}

Most theoretical studies of Bose-Einstein condensates (BEC)
in gases of alkali atoms confined in magnetic or optical traps 
have been conducted in the framework of the 
Gross-Pitaevskii (GP) equation \cite{gp}. 
The key point for the validity of this description is the
dilute condition of these systems, i.e., the average distance between
the atoms is much larger than the range of the inter-atomic interaction. In
this situation the physics is dominated by two-body collisions,
well described in terms of the $s$-wave scattering length
$a$.  The crucial parameter defining the condition for diluteness is the
gas parameter $x({\bf r})= n({\bf r}) a^3$, where $n({\bf r})$ is the
local density of the system. For low values of the average gas
parameter $x_{av}\le 10^{-3}$, the mean field Gross-Pitaevskii
equation does an excellent job (see for example 
Ref.~\cite{dal99} for a review). 
However, in recent
experiments, the local gas parameter may well exceed this value due to
the possibility of tuning the scattering length in the presence of a  
Feshbach resonance \cite{cornish00,goss00}. 

Under such circumstances it is unavoidable to test the accuracy of
the GP equation by performing microscopic calculations. If we consider cases
where the gas parameter has been driven to a region were one
can still have a universal regime, i.e., that the specific shape of
the potential is unimportant, we may attempt to describe the system as dilute
hard spheres whose diameter coincides with the scattering length.
However, the value of $x$ is such that the
calculation of the energy of the uniform hard-sphere Bose gas would
require to take into account the second term in the low-density
expansion \cite{fetter} of the energy density
\begin{equation}
  \frac {E}{V} = \frac {2 \pi n^2 a \hbar^2}{m}
  \left [ 1 + \frac {128}{15} \left ( \frac {n a^3}{\pi} \right)^{1/2}
          + \cdots \right ],
\label{low-ex}
\end{equation}
where $m$ is the mass of the atoms treated as hard spheres.
For the case of uniform systems, the validity of this expansion has been 
carefully studied using Diffusion Monte Carlo \cite{boro99} and
Hyper-Netted-Chain techniques \cite{mazz03}.

The energy functional associated with the GP theory is obtained
within the framework of  the local-density approximation (LDA) 
by keeping only the first
term in the low-density expansion of Eq.~(\ref{low-ex})
\begin{widetext}
\begin{equation}
  E_{\mathrm{GP}} [\Psi] = \int d{\bf r} \left [ \frac {\hbar^2 }{2m} \mid \nabla
  \Psi({\bf r}) \mid^2 + V_{\mathrm{trap}}({\bf r})\mid \Psi \mid ^2+ \frac {2 \pi
  \hbar^2 a }{m}\mid \Psi \mid ^4 \right ],
\label{func1}
\end{equation}
\end{widetext}
where 
\begin{equation}
  V_{\mathrm{trap}}({\bf r}) = \frac {1}{2} m (\omega_{\bot}^2 x^2 
  + \omega_{\bot}^2 y^2 +\omega_{z}^2 z^2 ) 
\label{trap}
\end{equation}
is the confining potential defined by the two angular frequencies
$\omega_{\bot}$ and $\omega_{z}$.
The condensate 
wave function $\Psi$ is
normalized to the total number of particles. 

By performing a functional variation of $E_{\mathrm{GP}}[\Psi]$ with respect
to $\Psi^*$ one finds the corresponding Euler-Lagrange equation, 
known as the Gross-Pitaevskii (GP) equation
\begin{equation}
  \left [ - \frac {\hbar^2}{2m} \nabla^2 + V_{\mathrm{trap}}({\bf r}) + 
   \frac{4\pi\hbar^2 a}{m} \mid \Psi \mid^2 \right ]\Psi=\mu\Psi , 
\label{gp1} 
\end{equation}
where $\mu$ is the chemical potential, which accounts for the conservation
of the number of particles. Within the 
LDA framework, the next step 
is to include into the energy functional of Eq.~(\ref{func1})
the next term of the low density expansion of Eq.~(\ref{low-ex}). 
The functional variation gives then rise to the so-called 
modified GP equation (MGP) \cite{fabro99} 
\begin{widetext}
\begin{equation}
  \left [ - \frac {\hbar^2}{2m} \nabla^2 + V_{\mathrm{trap}}({\bf r})+\frac {4 \pi \hbar^2 a}{m} \mid \Psi \mid^2 
    \left (1 + \frac {32 a^{3/2}}{3 \pi^{1/2}} \mid \Psi\mid \right)
    \right ] \Psi =  \mu \Psi .
\label{gp2}
\end{equation}
\end{widetext}
The MGP corrections have been estimated in  Ref.~\cite{fabro99} in a cylindrical 
condensate in the range of the scattering lengths and trap parameters
from the first JILA experiments with Feshbach resonances. These experiments took 
advantage of the  presence of a Feshbach resonance in the collision of two
$^{85}$Rb atoms to tune their scattering length \cite{cornish00}.
Fully microscopic calculations using  a hard-spheres interaction have
also been performed in the framework of Variational and Diffusion Monte
Carlo methods \cite{glyde1,glyde2,glyde3,blume1}. 

In this work we  compare the results of the GP and MGP equations
discussed above, Eqs.~(\ref{gp1}) and (\ref{gp2}),   
with variational Monte Carlo (VMC) calculations for
axially symmetric traps in a region ($x > 10^{-3}$) where the validity of the
GP equation is not clear. 
We examine both the 
ground state and excited states having a vortex line along
the $z$-axis. 

In the next Section we present 
our numerical approaches together with a discussion of ground state properties. 
In  Sect.~\ref{sec:vortex} we proceed to study several trial wave functions
to describe  the excited state with one vortex. A comparison between
VMC and the GP and MGP equations is done.  
We summarize our results in Sect.~\ref{sec:conclusions}.

\section{Numerical Approaches and Ground State Properties}\label{sec:groundstate}

The starting point of the Monte Carlo calculations is 
the Hamiltonian  for $N$ trapped 
interacting atoms  given by
\begin{equation}
  H = - \frac {\hbar^2}{2m} \sum_{i=1}^N \nabla_i^2 
  + \sum_{i=1}^N V_{\mathrm{trap}}({\bf r}_i) 
  + \sum_{i<j}^N V_{\mathrm{int}}(\mid{\bf r}_i -{\bf r}_j \mid ). 
\label{hamil}
\end{equation}
The  two-body interaction $V_{int}(\mid {\bf r}_i -{\bf r}_j \mid )$ between the
atoms is described  by  a hard-core potential of radius $a$, 
where $a$ is the scattering
length. The atoms are thus treated as hard spheres.
The next step is to define a trial wave function,
\begin{equation}
  \Psi_T (1, ...,N) = F(1,...,N) \Psi_{\mathrm{MF}}(1,...,N) 
\end{equation}
where $F(1,...,N)$ is a many-body correlation operator applied to the
mean-field wave function $\Psi_{\mathrm{MF}}$.  The advantage of using a
correlated trial wave function lies in the fact that non-perturbative
effects, as the short-range repulsion between atoms may be
directly incorporated into the trial wave function. The simplest
correlation operator has the Jastrow form \cite{jas55},
\begin{equation}
  F(1,...,N) = \prod_{i<j} f(r_{ij}).
\label{eq-jas}
\end{equation}
In our variational calculations we use a two-body correlation function 
which is the  solution of the Schr\"odinger equation for a pair of
atoms at very low energy interacting via a hard-core potential of
diameter $a$. The ansatz for the correlation function $f(r)$ reads 
\begin{equation}
  f(r) = \left\{
  \begin{array}{ll}
    \left ( 1 - a/r  \right )  & \,\,\,\,\, r>a \\
    0  & \,\,\,\,\, r \le  a \ .
  \end{array}
  \right.
\end{equation}
This type of correlation, besides being physically motivated, has been
successfully used in Refs.~\cite{glyde1,glyde2} to study both
spherically symmetric and deformed traps. These authors 
have also explored the quality of this correlation function by comparing Variational
Monte Carlo (VMC) and Diffusion Monte Carlo (DMC)
calculations for the case of spherically symmetric traps \cite{glyde3}, with a good
agreement between the VMC and DMC results.

The deformation of the trap is incorporated  in the mean field wave
function $\Psi_{MF}$, which is taken as the product of $N$  single
particle wave functions 
\begin{equation}
  \varphi({\bf r}) = A(\alpha)  \lambda^{1/4}
    \exp \left[-\frac{1}{2} \alpha (x^2 + y^2 + \lambda z^2 )\right] ,
\label{cond1}
\end{equation}
where $\alpha$ is taken as the variational parameter of the
calculation, and $A(\alpha)=(\alpha/\pi)^{3/4}$ is the  normalization constant. 
The parameter $\lambda = \omega_z/\omega_{\bot}$ is kept fixed and set
equal to the asymmetry of the trap. In this way the mean-field wave
function $\Psi_{\mathrm{MF}}$ has all the particles in the condensate, 
the latter being described by the wave function $\varphi$.

The evaluation of the expectation value of the Hamiltonian with
this correlated trial wave function provides an upper bound to the
ground state energy of the system
\begin{equation}
  E_T = \frac {\langle \Psi_T \mid H \mid \Psi_T \rangle}
              { \langle \Psi_T \mid \Psi_T \rangle }. 
\end{equation}
This expectation value has been evaluated by the Metropolis Monte Carlo 
method of integration
\cite{metro,guar}. 

The energy  obtained with the Hamiltonian of Eq.~(\ref{hamil})
can be directly compared with the output of the GP and MGP equations, see
Eqs.~(\ref{gp1}) and (\ref{gp2}).
The Gross-Pitaevskii equations represent a mean-field
description, with  all the atoms in the condensate. In fact, 
the additional correlations which are taken into account
 in the second order term of the low-density expansion of
the energy, see Eq.~(\ref{low-ex}), are incorporated in the 
density functional and, therefore, in the solution of the MGP equation.
In contrast,
the Monte Carlo calculation incorporates explicitly the inter-atomic
correlations and therefore one could in principle  find the natural
orbits and extract the occupation of the condensate \cite{glyde1}.

The GP and the MGP equations have been solved by the steepest descent 
method \cite{flo80} for the deformed harmonic oscillator trap previously 
described in Eq.~(\ref{trap}). 
An initial deformed trial state is projected onto the minimum of the 
functional by propagating it in imaginary time. In practice, one chooses 
a small time step $\Delta t$ and iterates the equation 
\begin{equation}
\Psi({\bf r},t+\Delta t) \approx \Psi({\bf r},t) - \Delta t H \Psi({\bf r},t)
\end{equation}
by normalizing $\Psi$  at each iteration. When the gas parameter becomes 
large, the time step which governs the rate of convergence should be taken 
accordingly small. 
Convergence is reached when the chemical potential becomes a constant
independent of the position, see Eqs.~(\ref{gp1}) and (\ref{gp2}).

For the comparison of the results obtained with the different GP type
equations and the variational Monte Carlo calculations, we consider a 
disk-shaped   trap with $\lambda = \omega_z/\omega_{\perp} = {\sqrt {8}}$,
 see  Ref.~\cite{cornell95}. 
We have fixed the scattering length
to  $a= 35 a_{\mathrm{Rb}}$, with $a_{\mathrm{Rb}}=100 a_0$, 
$a_0$ being the Bohr radius.
We set the number of confined atoms to $N=500$ in order 
to keep the amount of computing time acceptable when using the Monte
Carlo method.  All the numerical results are given in units
of the harmonic oscillator length 
$a_{\perp}= (\hbar / m \omega_{\perp})^{1/2}$ and the harmonic
oscillator energy $\hbar \omega_{\perp}$.
 
First we analyze the GP and the MGP results reported in
Table \ref{tab:1}. For a scattering length $a= 35 a_{\mathrm{Rb}}$, the corrections of
the MGP approach to the chemical potential are of the order of $20\%$.
The energy corrections  are also relevant and it is interesting to study the 
different contributions to  the energy. The kinetic energy is given by
\begin{equation}
  E_{\mathrm{kin}} = \frac {\hbar^2}{2m} 
  \int d {\bf r} \mid \nabla \Psi({\bf r}) \mid ^2 ,
\end{equation}
while the harmonic oscillator energy due to the
trapping potential reads
\begin{equation}
  E_{\mathrm{HO}} =\frac {m}{2} \int d {\bf r} ~  (\omega_{\perp}^2 (x^2 +
  y^2) + \omega_z^2 z^2)  \mid \Psi({\bf r}) \mid^2 ,
\end{equation}
and the interaction energies $E_1$ and $E_2$ are given by
\begin{eqnarray}
E_1 &=& \frac{2\pi\hbar^2 a}{m} \int d {\bf r}\mid\Psi({\bf r})\mid^4
  , \\
E_2 &=& \frac {2 \pi \hbar^2 a}{m} \frac {128}{15} 
  \left ( \frac {a^3}{\pi} \right )^{1/2} 
  \int d {\bf r} \mid \Psi({\bf r}) \mid ^5 .
\end{eqnarray}
The virial theorem is used to establish a relation between the
different contributions to the energy, viz.,
\begin{equation}
  2 E_{\mathrm{kin}} - 2 E_{\mathrm{HO}} + 3 E_1 + \frac {9}{2} E_2 = 0 ,
\end{equation}
which   serves as a proof of the numerical accuracy of the solution of
the GP equations. The results in Table~\ref{tab:1} show 
that this test is well satisfied by all calculations.

Notice that the kinetic energy associated with the mean field
descriptions is not negligible, indicating that the regime where the
Thomas-Fermi approximation to the GP equation is valid has not been
reached. In  this limit, the chemical potential is  
\begin{equation}
  \mu_{\mathrm{TF}} = \frac {1}{2} 
  ( 15 \bar {a} N \lambda)^{2/5} \hbar \omega_{\perp}
\end{equation}
where $\bar a= a/a_{\perp}$ is the dimensionless scattering length,
and the energy per particle $E_{\mathrm{TF}}/N = 5 \mu_{\mathrm{TF}} /7$. 
In this approach we have
$E_{\mathrm{TF}}/N =9.03 \hbar \omega_{\perp}$ and $\mu_{\mathrm{TF}}=12.64 
\hbar \omega_{\perp}$. 
Both these values differ from the 
values reported in Table~\ref{tab:1}. However, this approximation can still 
be used to estimate  the peak value of the gas parameter, namely
\begin{equation}
  x_{\mathrm{TF}}^{pk} =  n(0)a^3 = 
  \frac {1}{8 \pi} ( 15 \bar a N \lambda ) ^{2/5} \bar a^2 ,
\end{equation}
which yields
$x_{\mathrm{TF}}^{pk}=0.023$. At this rather large value of 
the diluteness parameter, the   corrections brought
by the MGP equation to the GP results are expected to be relevant \cite{boro99,
mazz03,fabro01}. 
However,  $x$ is low enough to allow for a  mean-field approach
(as it is the case of the MGP equation). For such density regimes,
a mean-field approach provides
a rather good description when
compared with a microscopic calculation~\cite{fabro99}.
 
The variational Monte Carlo results are also given in Table
\ref{tab:1} and they show a  close agreement with the
results provided by the MGP equation. Notice that in this approach,
and using the Hamiltonian of Eq.~(\ref{hamil}), the potential energy is
zero since the wave function is strictly zero inside the core.  The
total energy in this case is distributed between $E_{\mathrm{HO}}$ and the true
kinetic energy. Actually the only energies that can be directly
compared with the GP results  are the total and the harmonic
oscillator energies. 

The Monte Carlo results obtained with the Metropolis algorithm take into account the
energy of 27000 configurations, grouped in 90 blocks of 300 movements.
 At each Monte Carlo step we move all the particles
and the acceptance is around 58 $\%$. A thermalization process is incorporated 
at the beginning of the Monte Carlo process and before  each block.
In the Monte Carlo calculation we have used the
Pandharipande-Bethe prescription for the kinetic energy~\cite{guar}, 
which produces a smaller variance. 
To a give a feeling for the numerical accuracy of 
our VMC results, we list here GP, MGP and VMC results in the dilute limit.
We employ $N=500$ particles and a scattering length for $^{87}$Rb considered by 
Dalfovo and Stringari in Ref.~\cite{dalstr96},  which in units of 
the oscillator parameter 
perpendicular to the $z$-axis is  $4.33\times 10^{-3} a_{\perp}$.
We obtain energies in units of the oscillator energy of 3.3032, 3.3080
and 3.3242(1) for GP, MGP and VMC calculations, respectively.
The VMC results are for an optimum variational parameter $\alpha= 0.475$. 
Taking into account
that the two-body correlation has been kept fixed, and that the
only variational parameter is $\alpha$, these results indicates that our ansatz for the 
variational wave function is a viable one. Actually, as the reader will notice 
from the discussion 
below,
this discrepancy of roughly $0.5\%$ is
of the same relative order as for the higher density cases reported here.

\begin{figure}[hbtp]
\centering
\includegraphics[scale=0.5,angle=-90]{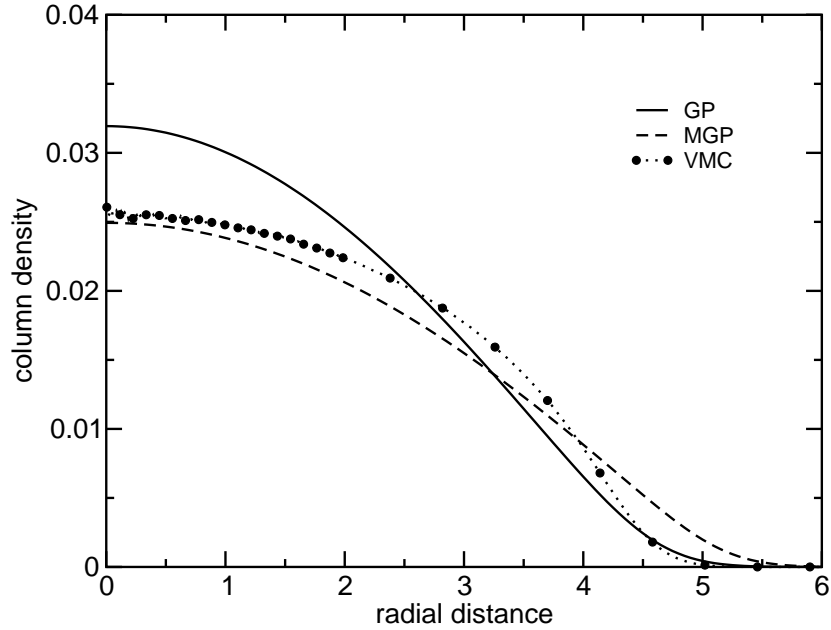}
\caption{Ground state column density $n_c({ r}_{\perp})$ 
as a function of the distance to the z axis, for
$N=500$ particles, comparing the GP (solid line) and MGP (dashed line) 
results for 
$a = 35 a_{\mathrm{Rb}} = 0.15155 a_{\perp}$. Also shown are the results of 
variational Monte Carlo calculations (line with symbols). 
The deformation $\lambda=\sqrt 8$ and the oscillator lengths are defined 
as in Refs.~\cite{dalstr96,cornell95}. The radial distance is given in 
units of $a_{\perp}=(\hbar / m \omega_{\perp})^{1/2}$. 
The column density is dimensionless. See text for further details.}
\label{fig:1}
\end{figure}

In the minimization process we keep fixed the parameter $\lambda$ in the
single particle wave function of Eq.~(\ref{cond1}), i.e., we assume that the
deformation of the trap is transferred to the wave function, and vary
only  $\alpha$. At the minimum, $\alpha = 0.7687 $. 
One can also explore the effects of the correlations in the density
profiles. These profiles, which represent a column density  defined according to
\begin{equation}
n_c(r_{\perp}) = \int dz~n(r_{\perp},z) 
\label{eq:profi}
\end{equation}
and normalized such that $ 2 \pi \int dr_{\perp}~r_{\perp}~ n_c(r_{\perp}) = 1 $,
 are shown in Fig.~\ref{fig:1} for the various
approximations used in this work. The repulsive character of the
correction term of the MGP equation translates into a decrease of the
value of the column density at the origin and an increase of the
size of the condensate \cite{fabro99,fabro01}.    
 This gives a slightly more extended profile for the
MGP approach compared with both the GP and the VMC results.
As one can see from Fig.~\ref{fig:1}, there is a much better agreement
between the Monte Carlo and MGP profiles than with the corresponding profile
from the GP calculation, particularly at small values 
of the radial distance where the density is 
larger.

The  good agreement between VMC and MGP does not guarantee
that these methods give a good description of the system. However, as it
was shown in Ref.~\cite{glyde2} for the case of spherical traps, the
improvements introduced in the trial wave function by a Diffusion Monte Carlo
calculation, which in principle allows for an exact solution of the many-body problem,
are rather small and the variational wave function of Eq.~(\ref{cond1})
provides a very good description of the system. Therefore we 
assume that the same will be true for deformed traps. 
Furthermore, for these values  
 of the diluteness parameter, 
the MGP  equation is very useful to calculate the energy,
chemical potential  and the density profiles of the ground state of
the system  for condensates with larger number of particles, which 
would be computationally prohibitive for a Monte Carlo calculation.

\section{Vortex  States}\label{sec:vortex}

The existence of these
excited condensate states is crucial to studies of  the superfluid behavior of
trapped atomic condensates.
In this section we study the effects  of correlations in 
vortex states. We consider a singly quantized vortex line 
 along the $z$-axis. This means that all the
atoms rotate around $z$-axis  with  a quantized circulation. 
 The GP equation can easily be generalized to
describe this kind of vortex states \cite{dal99} 
by using the   following ansatz for 
the condensate wave function 
\begin{equation}
\Psi({\bf r}) = \psi({\bf r}) \exp [i \kappa \phi] ,
\label{vor-1}
\end{equation}
where $\phi$ is the angle around the $z$-axis and $\kappa$ 
is an integer. This  vortex state has a tangential velocity
\begin{equation}
v_{\phi} = \frac {\hbar}{m r_{\perp}} \kappa
\end{equation}
where $r_{\perp}=\sqrt{x^2+y^2}$ is the distance to the symmetry
axis of the vortex.
The number $\kappa$ represents the quantum of circulation, 
and  the total angular
momentum along the $z$ axis is given by $N \kappa \hbar$. 
Introducing the wave function of Eq.~(\ref{vor-1}) in the GP energy functional 
of Eq.~(\ref{func1}) one gets the corresponding GP energy functional for the 
vortex state
\begin{widetext}
\begin{equation}
  E_{\mathrm{GP+vor}} [\Psi] = 
   \int d{\bf r} \left [ \frac {\hbar^2 }{2m} |\nabla \psi({\bf r}) |^2 
    + \frac {\hbar^2}{2m} \frac {\kappa ^2}{r_{\perp}^2} |\psi|^2 
    + V_{\mathrm{trap}}({\bf r}) |\psi|^2 
    + \frac {2 \pi \hbar^2 a }{m} |\psi|^4 \right ],
\label{func2}
\end{equation}
\end{widetext}
which incorporates a  centrifugal term in the density functional, arising from the
quantized flow of atoms around the vortex core. 
This term defines a rotational energy 
\begin{equation}
E_{\mathrm{rot}} = 
\frac {\hbar^2}{2m} \int d{\bf r }\,\frac{\kappa^2}{r_{\perp}^2}|\psi({\bf r})|^2.
\end{equation}

The corresponding nonlinear Schr\"odinger equation obtained 
by functional variation is
\begin{equation}
\left [ - \frac {\hbar^2}{2m} \nabla^2 
+\frac {\hbar^2}{2m} \frac {\kappa^2}{r_{\perp}^2}
+ V_{\mathrm{trap}}({\bf r}) 
+ \frac {4 \pi \hbar^2 a}{m} |\psi|^2 \right ] \psi = \mu \psi   .
\label{vor3}
\end{equation}
 
Adding $E_2$ to the density functional and after performing a functional 
variation one gets the corresponding MGP equation for the vortex state.
 
Based on the virial theorem, one can again derive a relation between the
different contributions to the energy:
\begin{equation}
  2 E_{\mathrm{kin}} - 2 E_{\mathrm{HO}} + 3 E_1 + \frac {9}{2} E_2 + 2 E_{\mathrm{rot}}= 0 .
\end{equation}
 
The thermodynamic   critical angular frequency $\Omega_c$ required to 
produce a vortex of vorticity $\kappa$ is obtained by comparing 
the energy of the system in the rotating frame with and without the vortex \cite{fetter2001}:
\begin{equation}
 \Omega_c = 
\frac {1}{N \hbar \kappa} \left [  E_{\kappa} - E_0 \right ] .
\end{equation}
 
A main feature of a vortex state is  the hole (core of the vortex) that appears in the center 
of the density profile along the rotation axis. From Eq. (\ref{vor3}), it is clear  that
the solution of this equation has to vanish on the $z$-axis due to the presence  of the
centrifugal term.
The size of the core is characterized by the healing length.
%Simple
%estimations \cite{fetter2001} of $\xi$ and $\Omega_c$ are  
%\begin{equation}
%\xi = (8 \pi n(0) a )^{-1/2}
%\label{eq:xi} 
%\end{equation}
%and 
%\begin{equation}
%\Omega_c = ..... .   
%\label{eq:ome}
%\end{equation}
 
For the microscopic description of the vortex state we use an  
Onsager-Feynman type  trial wave function \cite{fey}
\begin{equation}
\Psi_F (1,...,N) = e^{i \kappa \sum_j \phi_j} 
\prod_j f(r_{\perp ,j}) \Psi_0(1,...,N),
\label{eq:wf0}
\end{equation}
where $\Psi_0(1,...,N)$ is the ground state wave function. 
The phase factor $\kappa\sum_j \phi_j$ depends on the angular variables 
of the particles and is the equivalent to  the phase factor introduced in
the mean field description  of Eq.~(\ref{vor-1}).
The  function $f(r_{\perp})$ modulates the density as a function of
the radial coordinate $r_{\perp}$. 
We examine three types of $f(r_{\perp})$. 
In the first ansatz we use the simple option
\begin{equation}
f_1(r_{\perp})= r_{\perp} . 
\label{eq:wf1}
\end{equation}
In the second case we consider,
\begin{equation}
f_2(r_{\perp})= 1 -\exp\left ( {- r_{\perp}/d }\right ),
\label{eq:wf2}
\end{equation}
where $d$ is a variational parameter.
Notice that for $d=1$, the
behavior of $f_2(r_{\perp})$ for small $r_{\perp}$ coincides with
the behavior of $f_1(r_{\perp})$. 
Finally the third function,  is that of Ref. \cite{reatto}
which has been used in the context of quantum liquids,
\begin{equation}
f_3(r_{\perp}) = 1 - \exp \left ( {- (r_{\perp}/d)^2 }\right ),
\label{eq:wf3}
\end{equation}  
where $d$ is  again a variational parameter. 

These three trial wave functions 
describe a singly quantized vortex 
state ($\kappa =1$), whose axis lies in the $z$-direction and with a tangential 
velocity field $v_{\phi} = \hbar/m r_{\perp}$.  
The evaluation of  the expectation value of the Hamiltonian 
(Eq.~(\ref{hamil})) with these wave functions is equivalent 
to calculate the mean value of the Hamiltonian
\begin{widetext}
\begin{equation}
  H = - \frac {\hbar^2}{2m} \sum_{i=1}^N \nabla_i^2 
  + \sum_{i=1}^N \frac {\hbar^2}{2m}  \frac {\kappa^2}{r_{\perp,i}^2}
  + \sum_{i=1}^N V_{\mathrm{trap}}({\bf r}_i) 
  + \sum_{i<j}^N V_{\mathrm{int}}(\mid{\bf r}_i -{\bf r}_j \mid ) , 
\label{hamil-vor}
\end{equation}
\end{widetext}
with $\Psi(1,...N) = \prod_j f(r_{\perp,j}) \Psi_0(1,...,N) $.
In this way the
rotational contribution to the energy has been directly incorporated 
in the Hamiltonian. Minimizing this new
problem provides the best energy and wave 
functions inside this subspace of wave functions.
In the context of liquid $^4$He there have been attempts to perform a 
full minimization allowing for a more general phase function. 
The analysis indicates that the present procedure provides
very accurate results \cite{boro2}.
 
We start by discussing the GP and MGP results
(obtained by the steepest descent method \cite{flo80} as done  
for the ground state as well) with  an
initial condensate wave function
\begin{equation}
\psi ({\bf r}) \propto f_1(r_{\perp}) \Psi_0({\bf r}).
\end{equation}
It is worth mentioning, as a check of the numerical procedure,
 that starting with  $f_2(r_{\perp})$ or $f_3(r_{\perp})$
to modulate the condensate 
wave function we converge to  the same results as with $f_1(r_{\perp})$.

 As expected, the presence 
of the vortex increases  the chemical potential. 
Also $E_{\mathrm{HO}}$ has a small increase, related to the enlargement of the
profile due to the presence of the vortex hole.
These results are listed in Table \ref{tab:2}.
Although the MGP corrections to the energy are sizable and of the same 
order as those in the ground state, the critical frequency, 
$\Omega_{GP}=0.29 \omega_{\perp}$ , is barely
affected as both energies, the energy of the vortex state and the ground state 
energy, are shifted by similar amounts, 
yielding  $\Omega_{MGP}= 0.24 \omega_{\perp}$.

\begin{figure}[hbtp]
\centering
\includegraphics[scale=0.5,angle=-90]{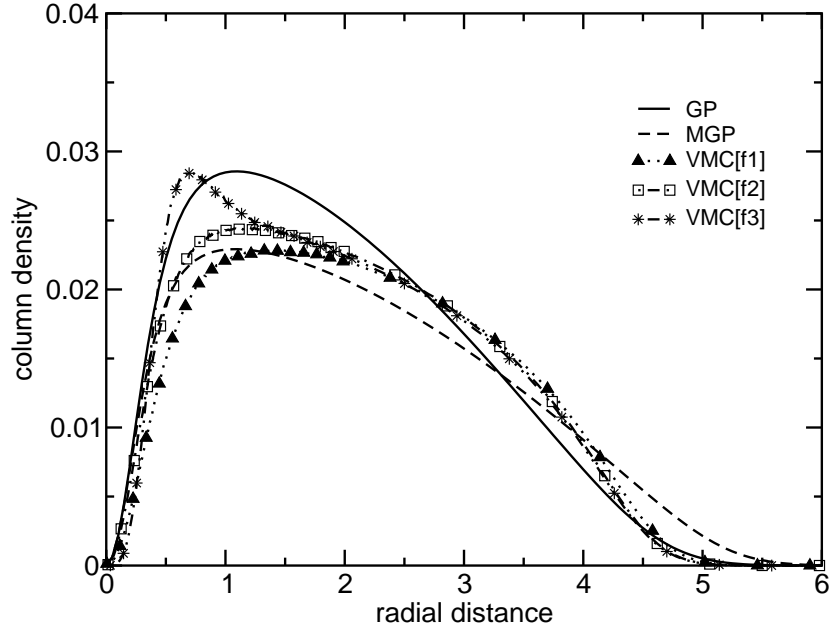}
\caption{Vortex column density $n_c({\bf r}_{\perp})$ 
as a function of the distance to the z  axis, for
$N=500$ particles, comparing GP (solid line) and MGP (dashed line) results for 
$a = 35 a_{\mathrm{Rb}} = 0.15155 a_{\perp}$. Also shown are the results of 
variational Monte Carlo calculations (lines with symbols) using the
different
Onsager-Feynman ansatzes. 
The deformation $\lambda=\sqrt 8$ and the oscillator lengths are defined 
as in Refs.~\cite{dalstr96,cornell95}. The radial distance is given in
units of $a_{\perp}=(\hbar / m \omega_{\perp})^{1/2}$. 
The column density is dimensionless. See text for further details.} 
\label{fig:2}
\end{figure}

The  GP and MGP profiles for the vortex state are shown in Fig. ~\ref{fig:2}. 
As a consequence of the repulsive character of the MGP 
corrections, the central density of the 
GP ground-state density profile is higher than the MGP one and, therefore,
the depth of the hole around the $z$ axis is larger in the GP
approach. However, the healing length is almost the same.

%One can qualitatively understand these results 
%by  recalling that   
%that the MGP equation, implies
%a more repulsive effective interaction than in the GP case. In some sense, the MGP approach
%produces similar results in energy and density profiles than the GP equation, if in the
%GP one uses an  
% appropriate larger scattering length. 
%We see  in the simple expresions ( Eqs. (\ref{eq:xi}) (\ref{eq:ome})), that 
%there is a compensation between the increase in the scattering length and the 
%decrease of the central density in such a way that the variation of $\Omega_c$ and $\xi$ is 
%very small. 

As can be seen from Table \ref{tab:3}, the 
Monte Carlo results for the energies are in  good 
agreement with the MGP ones for all the trial  wave functions considered.
This table shows two types of calculations. In the first three rows we list the 
energies obtained by keeping 
$\Psi_0$ equal to the ground state wave function
and performing the minimization with respect to the parameter $d$ 
in the modulating function,
except in the case of $f_1$, which has no variational parameters.
In the second set of results, we  perform a minimization allowing to vary also the 
harmonic oscillator parameter $\alpha$ of the wave function $\Psi_0$. The changes in 
$\alpha$ and $d$ do not yield significant  changes in the computed energy. 

The density profiles seem to be more sensitive to the modulating function,
as one can see from Fig.~\ref{fig:2} . These profiles correspond to the case where
the ground state wave function $\Psi_0$ is kept fixed when we minimize the energy
of the vortex state. For $f_3$ we obtain a radial structure which is not present
in the mean field approach \cite{reatto}.
The MGP profile shows a broader surface region than the VMC profiles. In the core of 
the vortex, the MGP profile looks very similar to the VMC results with the 
modulating function $f_2$ of Eq.~(\ref{eq:wf2}). 
These two results exhibit a smaller healing length than the 
VMC calculation which employs $f_1$.

From the variational point of view, the best description of the vortex should correspond to 
the wave function that provides the minimum energy. According to this criterion, this 
corresponds to the trial wave function built with the modulating function $f_1$ of Eq.~(\ref{eq:wf1}).

\begin{figure}[hbtp]
\centering
\includegraphics[scale=0.5,angle=-90]{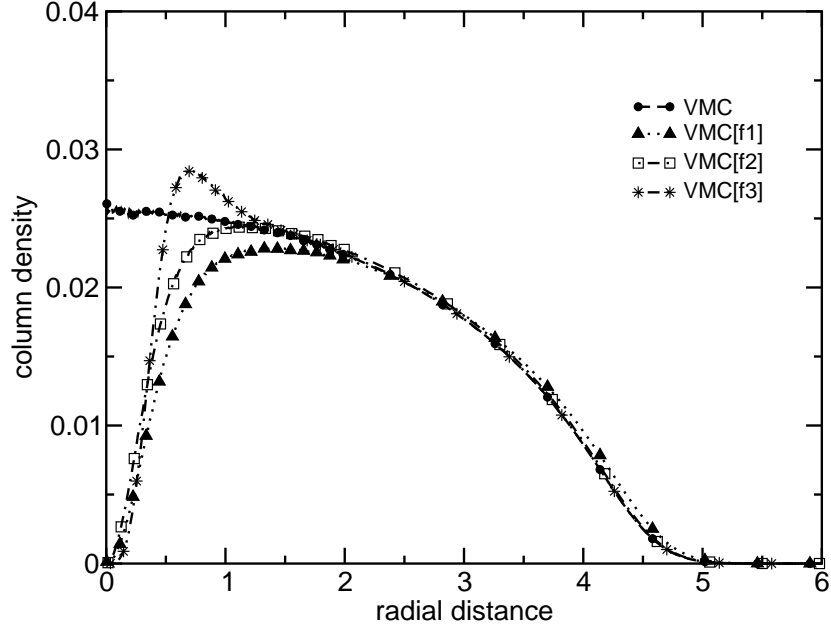}
\caption{Column density $n_c({\bf r}_{\perp})$ 
as a function of the distance to the z  axis, 
comparing the VMC profiles for the vortex
state corresponding to the different Onsager-Feynman ansatzes (lines with symbols
as in Fig.(\ref{fig:2}) and the ground
state (dashed line with full circles). 
The trap parameters and the scattering length are the same 
as in the two preceeding figures. 
The radial distance is given in
units of $a_{\perp}=(\hbar / m \omega_{\perp})^{1/2}$.
The column density is dimensionless. 
See text for further details.}
\label{fig:3}
\end{figure}

Finally, in Fig.~\ref{fig:3} we plot the density profiles for all VMC calculations,
with and without vortices.  
We note that they all provide a similar healing length  
and that the asymptotic behavior is almost equal for both the ground state
and the vortex states.

\section{Conclusions}\label{sec:conclusions}

We have compared the results of the Gross-Pitaevskii (GP) and the modified 
Gross-Pitaevskii (MGP) equations with ab initio variational
Monte Carlo calculations for Bose-Einstein condensates of atoms in  deformed traps.
We have studied both the ground-state and excited states having a vortex line along the
$z$-axis. The interatomic potential has been characterized by a hard-sphere potential
with a  radius which 
coincides with the scattering length used in the GP and MGP equations.

We have performed the calculations for 500 particles. The parameters
characterizing the trap and the scattering length have been chosen to reach 
values of the gas parameter where the 
MGP calculations provide corrections of the order of $20 \%$ compared with the 
GP results. It is indeed very interesting that even at 
such values of the gas parameter one can 
still describe the system in terms of mean-field approaches.
We find for example an excellent agreement between the MGP and the VMC results,
especially for the energies of the ground state and the 
vortex states. The MGP and VMC density profiles for the ground state
are also in good agreement.
The situation is different for the vortex state. Three different trial 
wave functions produce similar energies but slightly  different profiles. 
In the core of the vortex, the MGP profile is close to the 
profiles obtained with the ansatzes $f_1$ and $f_2$ of Eqs.~(\ref{eq:wf1}) and (\ref{eq:wf2}),
respectively. The latter two wave functions yield also the lowest energies.
Whether a Diffusion Monte Carlo (DMC) 
calculation will show a similar trend remains to see. We are planning
DMC studies of the systems discussed here.
Our preliminary DMC calculations for the energy
of the ground state
show little change with respect to the VMC results and, hence, a very
 good agreement with the MGP results. 

In summary, we would like to point out that the good
agreement between the VMC and MGP is rather encouraging, and allows for further MGP
explorations of vortex states in condensates with both a larger number 
of interacting atoms and  large scattering lengths. 

\section*{Acknowledgments}

The authors are grateful to Professors A.~Fabrocini and J.~Boronat for
many useful discussions. This research was also partially supported by
DGICYT (Spain) Project No.~BFM2002-01868 and from Generalitat de
Catalunya Project No.~2001SGR00064. J.~Mur-Petit acknowledges support
from the Generalitat de Catalunya. Support from the Research Council of Norway
is acknowledged.

\newpage

\clearpage

\begin{table}[hbtp]
\caption{Chemical potential and energies in units of $\hbar \omega_{\perp}$ 
from the GP, MGP and VMC
calculations for the ground state. The scattering length is 
$a= 35 a_{\mathrm{Rb}}=0.15155 a_{\perp}$,
$\lambda=\sqrt{8}$, $N=500$.}
 
\begin{ruledtabular}
\begin{tabular}{ccccccc}
      & $\mu$  & $E/N$  & $E_{\mathrm{kin}}/N$& $E_{\mathrm{HO}}/N$ & $E_{1}/N$& $E_{2}/N$ \\\hline
GP    & 12.980 & 9.496836~~~  & 0.39495~~~  & 5.61911~~~  & 3.4827765  & --       \\
MGP   & 15.453 & 11.06108~~~  & 0.35353~~~  & 6.94092~~~  & 2.516691   & 1.249938 \\
VMC   &   --   & 11.12109(14) & 4.21520(24) & 6.90590(19) &    --      & --       \\
\end{tabular}
\end{ruledtabular}
\label{tab:1}
\end{table}

\begin{table}[hbtp]
\caption{Chemical potential and energies in units of $\hbar \omega_{\perp}$
from the GP and MGP 
calculations for the one-vortex state with the vortex line along the
$z$-axis.
The scattering length is 
$a= 35 a_{\mathrm{Rb}}=0.15155 a_{\perp}$, $\lambda=\sqrt{8}$, $N=500$.}

\begin{ruledtabular}
\begin{tabular}{cccccccc}
& $\mu$  & $E/N$    & $E_{\mathrm{kin}}/N$& $E_{\mathrm{HO}}/N$ & $E_{1}/N$&
$E_{2}/N$& $E_{\mathrm{rot}}/N$ \\ \hline
GP-1v & 13.187 & 9.7835936 & 0.42508   & 5.74271  & 3.403871  & -- & 0.21193 \\
MGP-1v& 15.623 & 11.305   & 0.37692    & 7.03774  & 2.482418  &
1.223280 & 0.18492 \\
\end{tabular}
\end{ruledtabular}
\label{tab:2}
\end{table}

\begin{table}[hbtp]
\caption{Variational Monte Carlo results obtained with different 
Onsager-Feynman ansatzes.
 The results labeled VMC[f1], VMC[f2]
and VMC[f3] stand 
for the modulating wave functions in Eqs.~(\ref{eq:wf0}), (\ref{eq:wf2}), 
and (\ref{eq:wf3}) respectively.}
\begin{ruledtabular}
\begin{tabular}{ccccccc}
& $\alpha$ &  d &    $E/N$ &  $E_{\mathrm{kin}}/N$ &  $E_{\mathrm{HO}}/N$ & $ E_{\mathrm{rot}}$ \\ \hline

VMC[f1]&  0.7685 & --    & 11.33432(18) & 4.14804(26) & 7.02175(19) & 0.164527(44) \\
VMC[f2]&  0.7685 & 1.175 & 11.36273(18) & 4.23594(31) & 6.93987(24) & 0.186912(58) \\
VMC[f3]&  0.7685 & 0.425 & 11.39171(18) & 4.28845(30) & 6.91368(23) & 0.189580(30) \\ \hline

VMC[f1]&  0.775 & --    & 11.33415(17) & 4.18634(29)  & 6.98213(22) & 0.165679(60) \\
VMC[f2]&  0.745 & 1.425 & 11.35457(15) & 4.07816(31)  & 7.09696(25) & 0.179446(93) \\
VMC[f3]&  0.745 & 0.550 & 11.38683(19) & 4.14902(33)  & 7.06446(26) & 0.173350(26) \\

\end{tabular}
\end{ruledtabular}
\label{tab:3}
\end{table}
\end{document}